\documentclass[pra,aps,twocolumn]{revtex4}
\usepackage{amsmath,amssymb,graphicx,subfigure,appendix}
\begin{document}

\title{A proposed spin qubit CNOT gate robust against noisy coupling}
\author{J.~P. Kestner and S.~Das Sarma}
\affiliation{Condensed Matter Theory Center, Department of Physics, University of Maryland, College Park, MD 20742}

\begin{abstract}
We propose an implementation of the two-qubit gate in a quantum dot spin qubit system which is immune to charge noise problems.  Our proposed implementation, if it could be realized in a physical system, would have the advantage of being robust against uncertainties and fluctuations in the tunnel coupling and barrier gate voltage pulse area.  The key idea is to introduce an auxiliary dot and use an analog to the stimulated Raman adiabatic passage pulse sequence in three-level atomic systems, often referred to in the context of electron transport in quantum dot systems as Coherent Tunneling by Adiabatic Passage.  Spin-dependent tunneling opens the possibility of performing two-qubit gate operations by this method.
\end{abstract}

\maketitle
\section{Introduction}
We theoretically consider an alternative to the well-known Loss-DiVincenzo \citep{Loss98} implementation of the $\sqrt{\text{SWAP}}$ gate in a quantum dot spin qubit system.  The exchange-coupled gate has already been realized in experiments \citep{Petta05}, but it requires an exact knowledge of the coupling, $J$, in order to precisely control the gate voltage pulse area (i.e., the integral of the applied gate voltage over the pulse time).  In a scalable system, it is inevitable that the exchange coupling will vary spatially from one pair of dots to another, and precisely characterizing every link in a large quantum dot array is likely to be impractical.  Furthermore, even in the double dot system, temporal charge noise can affect $J$, resulting in gate errors \citep{Hu06,Culcer09}.  In fact, it is widely believed that charge noise is the currently limiting dephasing mechanism in the existing semiconductor quantum dot spin qubits because of the invariable presence of random charged traps and impurities in the environment that indirectly affect the exchange coupling $J$ through the microscopic quantum confinement potentials \citep{Nguyen11}.

Our proposed implementation has the advantage of being robust against uncertainties and fluctuations in the tunnel coupling and gate voltage pulse area.  The disadvantage of our approach is that it is slower and somewhat susceptible to noise in the interdot detuning.  However, we shall show that for low-frequency noise our proposed gate is sensitive only to the actual (rare) switching events and not the random nature of the quasi-static parameters, unlike the exchange-coupled gate.  This could markedly reduce the error rate to well below the quantum error correction threshold, without the need for an exact knowledge of $J$ across the sample or carefully tuning experimental parameters to a ``sweet spot" where $J$ is insensitive to noise sources \citep{Stopa08,Li10}, both daunting tasks in a large system.  Our proposal in its idealized abstraction is not specific to a particular physical system, and is in principle applicable to any system although the physical constraints on various energy and time scales (as discussed below) at this stage of materials development may limit its immediate implementation and usefulness.  Here we simply wish to show the general principle of the proposal, and how it is complementary to the exchange gate.  For a given system, a detailed analysis including all the specifics would be in order.

\section{Adiabatic protocol}
The central idea is to use an analog to the stimulated Raman adiabatic passage (STIRAP) pulse sequence in three-level atomic systems \citep{Bergmann98}.  This analogy has recently been invoked as a means of coherently transporting electrons between dots \citep{Greentree04,Cole08}, often referred to as CTAP (Coherent Tunneling by Adiabatic Passage).  Realizing spin-dependent tunneling opens the possibility of performing two-qubit gates by this method.

\begin{figure}
  \subfigure[]{\includegraphics[width=.28\columnwidth]{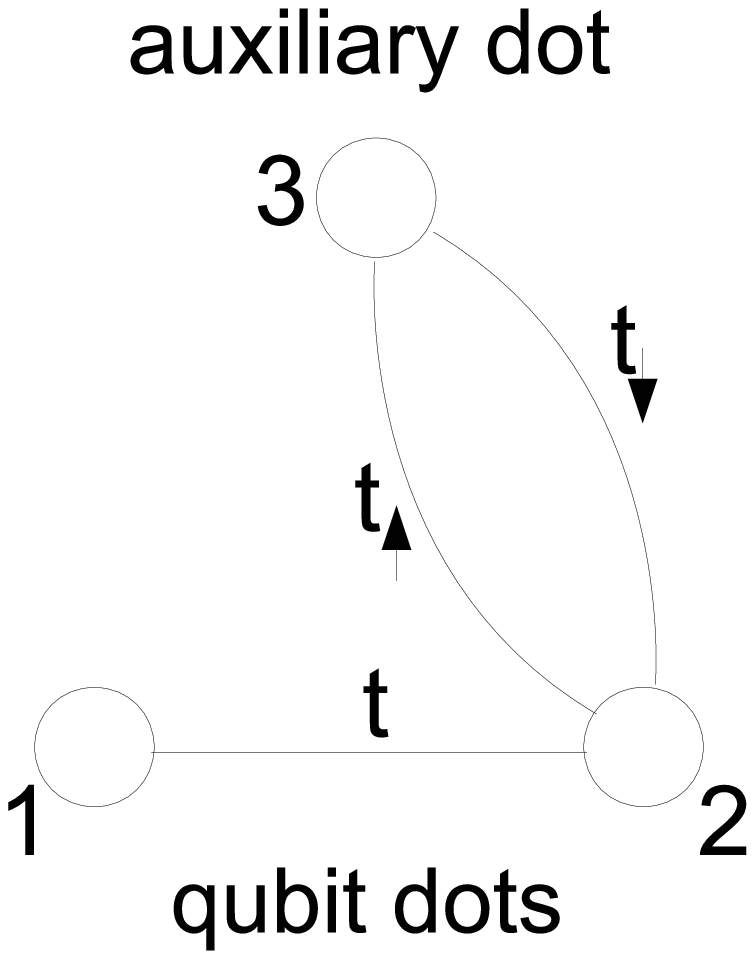}\label{fig:cnota}}
  \subfigure[]{\includegraphics[width=.7\columnwidth]{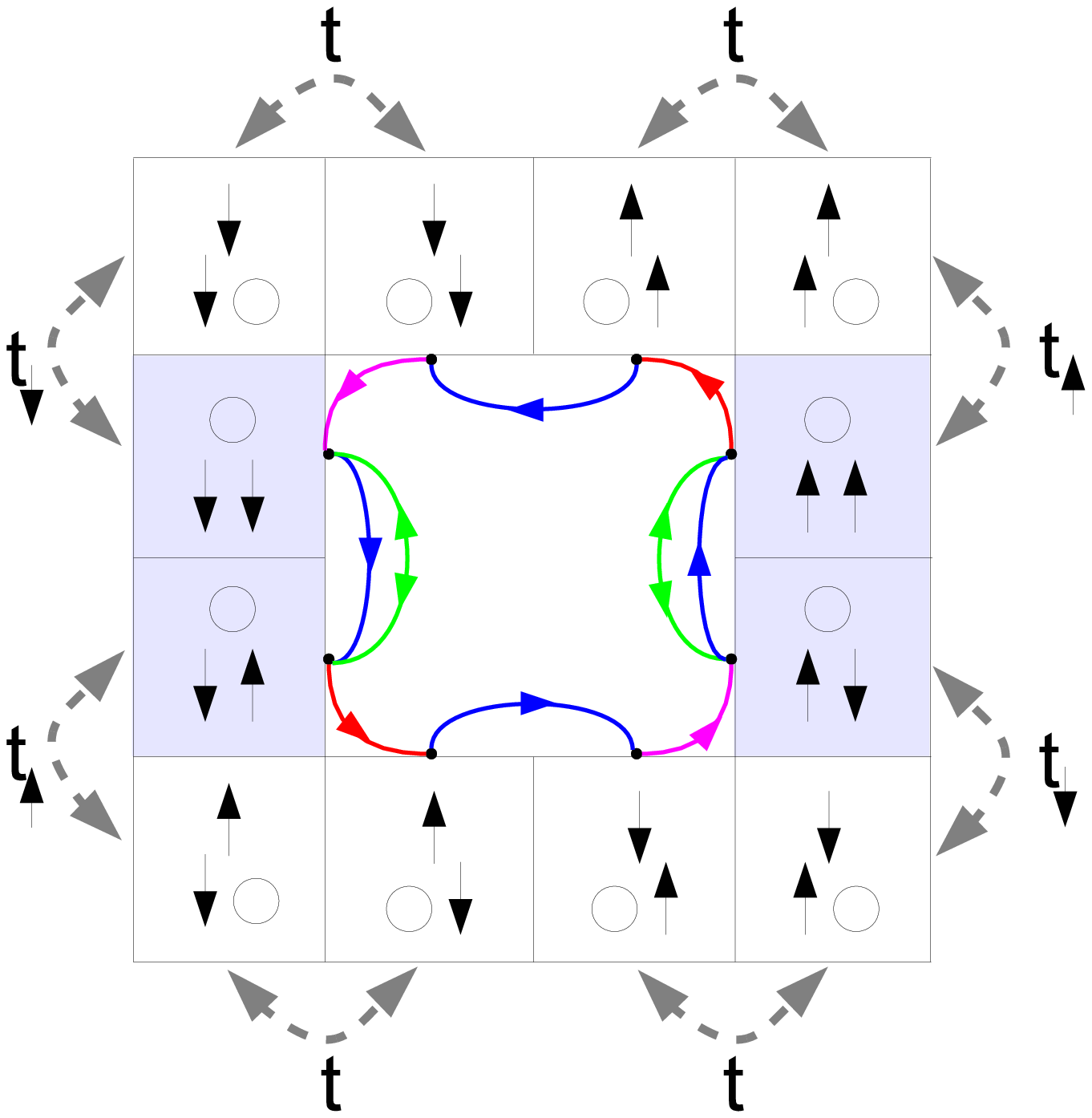}\label{fig:cnotb}}
  \caption{(Color online.) (a) The triple dot system with spin-dependent coupling to the auxiliary third dot. (b) Low-energy states of the triple dot.  The physical two-qubit space is shaded.  The directed curves inside the square correspond to transitions induced under the CNOT pulse sequence.  (See Fig.~\ref{fig:cnotc}.)}
\end{figure}

We consider two electrons in the triangular triple dot shown in Fig.~\ref{fig:cnota}.  Similar arrangements have previously been fabricated \citep{Gaudreau06,Schroer07,Laird10}.  (Although we could also use a linear triple dot, we shall see later that the triangular structure is preferable so that the direct intersite Coulomb interaction energy is the same regardless of which two dots are occupied and adiabaticity is more easily satisified.)  The qubit is defined as the spin state of a single occupied dot.  Thus, dot 3 must be unoccupied when the system is in the two-qubit subspace.
\begin{figure}
  \includegraphics[width=.8\columnwidth]{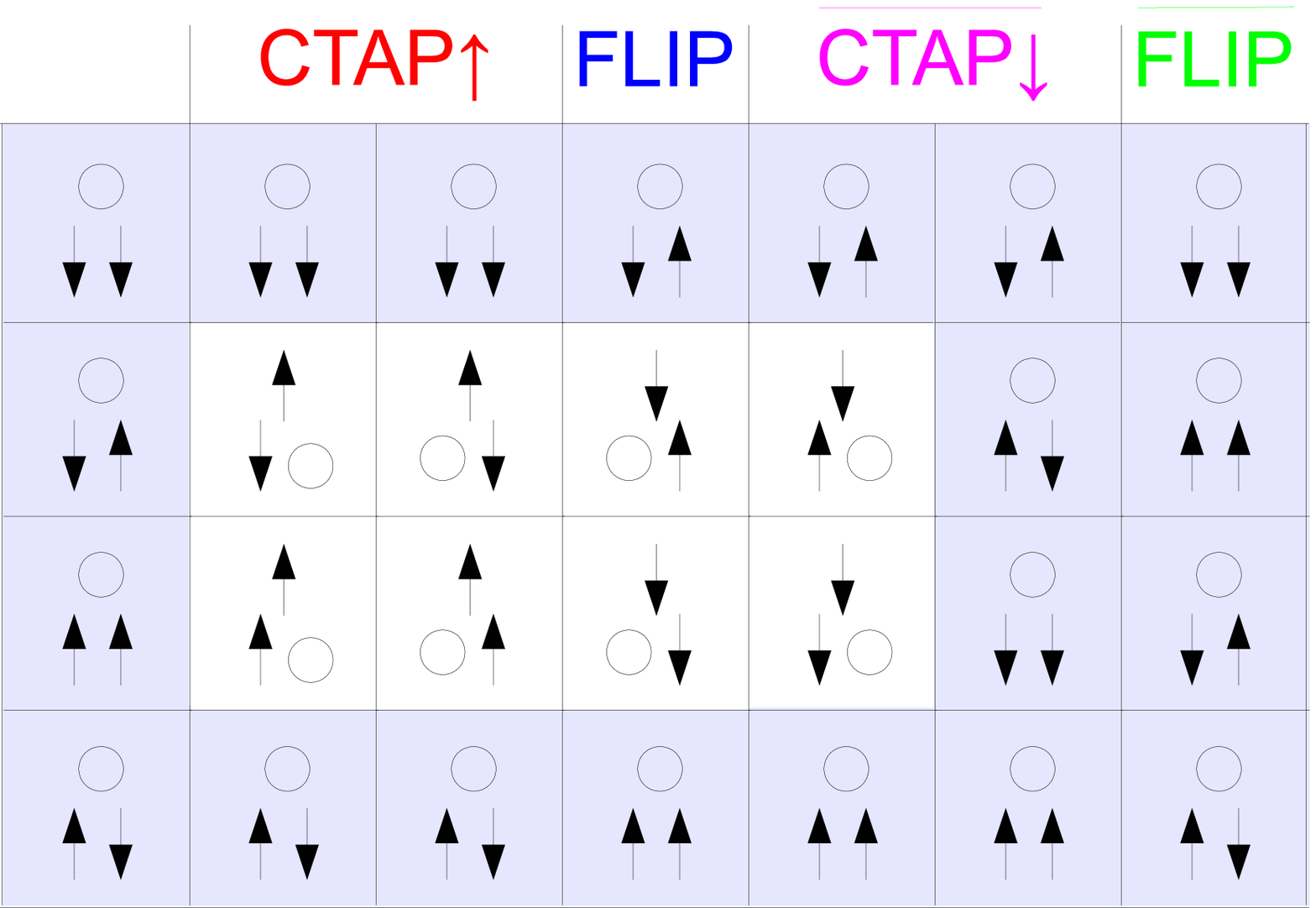}
  \caption{(Color online.) Depiction of the evolution of the physical two-qubit states under the CNOT pulse sequence.}\label{fig:cnotc}
\end{figure}

The two qubit dots are coupled via the usual spin-independent tunneling, which can be tuned via the barrier height by controlling the relevant gate voltage.  The auxiliary dot is coupled to one of the qubit dots via spin-dependent couplings which are controlled independently.  We assume that the excited levels in each dot are sufficiently high that we can ignore them and focus on the lowest orbital.  We also neglect spin-orbit couplings and other small corrections that would be accounted for in a thorough analysis of a specific system.  Then all remaining two-electron states and the couplings between them are depicted in Fig.~\ref{fig:cnotb}.

One can coherently transfer the state $\binom{0}{\sigma'\sigma}$ to $\binom{\sigma}{0\,\sigma'}$ by first turning on tunneling between the dots 1 and 2, then adiabatically turning off that tunneling while adiabatically turning on the spin-$\sigma$ tunneling to the auxiliary dot 3.  (The spin-dependent tunneling may afterwards be turned off on any timescale.)  We will refer to this as a $\text{CTAP}\sigma$ pulse sequence, and the reverse sequence as $\overline{\text{CTAP}\sigma}$.  By combining such a process with a spin flip operation on dots 2 and 3, denoted as FLIP (a $\pi$-pulse) and $\overline{\text{FLIP}}$ (a $3\pi$-pulse), we construct a CNOT gate in Fig.~\ref{fig:cnotc}.  Although we have shown the intermediate state of the CTAP processes in the table, we note that the intermediate state is not actually occupied during a sufficiently adiabatic process.  During this pulse sequence, the exact values of the couplings are irrelevant.  What is important is only that they change adiabatically.

Below we consider effects of fluctuations in tunnel couplings and detunings, finite ramping times, and finite superexchange on the procedure outlined above.  We then discuss the requirements for our proposal to be applicable to a physical system.  We show that for a particular example of low-frequency noise, the adiabatic CNOT gate offers a decrease of two orders of magnitude in the error rate compared to the standard exchange gate, sufficient to reach the quantum error correction threshold.

If we label the states from 1-12 starting at state $\binom{\downarrow}{0\,\downarrow}$ and going counterclockwise around Fig.~\ref{fig:cnotb}, the (time-dependent) model Hamiltonian is
\begin{equation}
H = \left(
  \begin{array}{cccccccccccc}
    \Delta_1 & t & 0 & 0 & 0 & 0 & 0 & 0 & 0 & 0 & 0 & F \\
    t & \Delta_2 & t_{\downarrow} & 0 & F & 0 & 0 & 0 & 0 & 0 & 0 & 0 \\
    0 & t_{\downarrow} & 0 & F & 0 & 0 & 0 & 0 & 0 & 0 & 0 & 0 \\
    0 & 0 & F & 0 & t_{\uparrow} & 0 & 0 & 0 & J & 0 & 0 & 0 \\
    0 & F & 0 & t_{\uparrow} & \Delta_2 & t & 0 & 0 & 0 & 0 & 0 & 0 \\
    0 & 0 & 0 & 0 & t & \Delta_1 & F & 0 & 0 & 0 & 0 & 0 \\
    0 & 0 & 0 & 0 & 0 & F & \Delta_1 & t & 0 & 0 & 0 & 0 \\
    0 & 0 & 0 & 0 & 0 & 0 & t & \Delta_2 & t_{\downarrow} & 0 & F & 0 \\
    0 & 0 & 0 & J & 0 & 0 & 0 & t_{\downarrow} & 0 & F & 0 & 0 \\
    0 & 0 & 0 & 0 & 0 & 0 & 0 & 0 & F & 0 & t_{\uparrow} & 0 \\
    0 & 0 & 0 & 0 & 0 & 0 & 0 & F & 0 & t_{\uparrow} & \Delta_2 & t \\
    F & 0 & 0 & 0 & 0 & 0 & 0 & 0 & 0 & 0 & t & \Delta_1 \\
  \end{array}
\right)
\end{equation}
where $t$ is the tunnel coupling between qubit dots, $t_{\sigma}$ is the spin-dependent tunnel coupling to the auxiliary dot 3, $\Delta_j = E_3-E_j$ is the detuning of the auxiliary dot 3 with respect to dot $j$, $F$ is the spin-flip pulse on dots 2 and 3, and $J$ is the superexchange.  All of these parameters (and their fluctuations) are in principle independent.  In the following, though, we will assume the usual relation $J \sim t^2/U$, with $U$ an on-site Coulomb interaction.  We use $\hbar = 1$ throughout and use the maximum tunnel coupling, $t_0$, as the unit of energy.

\begin{figure}
  \subfigure[]{\includegraphics[width=.98\columnwidth]{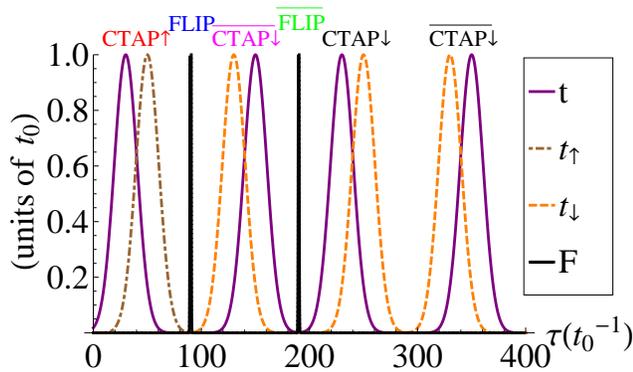}\label{fig:pulseseq}}
  \subfigure[]{\includegraphics[width=.98\columnwidth]{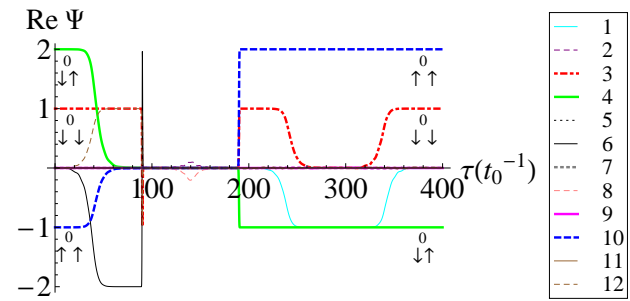}\label{fig:evolution}}\\
  \caption{(Color online.) (a) Pulse sequence used to perform CNOT in the numerical calculations. (b) Real part of the evolution of initial state $|\Psi\left(0\right)\rangle =|3\rangle + 2|4\rangle -|10\rangle$ to $|\Psi\left(400/t_0\right)\rangle =|3\rangle - |4\rangle +2 |10\rangle$ under the sequence (a) for $\Delta_i = 0$, $J=0$.}
\end{figure}

Here nonzero $J$ introduces unwanted swapping processes that result in gate errors, which we characterize via the error rate $\epsilon = \left|1-\langle \Psi\left(\tau\right)\left|\text{CNOT}\right| \Psi\left(0\right) \rangle / \langle \Psi\left(0\right)| \Psi\left(0\right) \rangle \right|$, where $\tau$ is the duration of the pulse sequence.  To avoid these errors we must have $J\tau \ll 1$, but on the other hand, adiabaticity requires $\tau\gg 1/t_0$.  So, since we assume $J \sim t^2/U$, we must have $U\ggg t_0$.  Using the pulse sequence of Fig.~\ref{fig:pulseseq} (we explain the extra pulses below) and averaging over initial states \cite{note}, the average error rate for $U=100t_0$ in the absence of noise is $\epsilon_{\text{avg}} \sim 10^{-2}$, and to reduce errors to the quantum error correction threshold, $\epsilon_{\text{avg}} \sim 10^{-4}$, requires $U/t_0 \geq 10^3$.  Thus the gate speed, set by $t_0$, is limited by the maximum attainable value of $U$.

By design, the adiabatic CNOT gate is insensitive to the shapes and areas of the tunnel coupling pulses, as long as adiabaticity is preserved.  This is true even when the pulses are jittery.  For instance, we have checked numerically that for noise in the tunneling rates from several random telegraph sources such that the standard deviation is 2\% and the average time between jumps, $T_{\text{RTN}}$, is an order of magnitude larger than the tunneling time, $\epsilon_{\text{avg}} \sim 10^{-4}$.  If the three tunneling rates fluctuate in lockstep rather than independently, or if $T_{\text{RTN}}$ is increased, sub-threshold error rates can be maintained even for fluctuations of 15\% or more.

However, even a very small constant value of $\Delta_1$ causes the control qubit to pick up a phase $\theta$ during the CNOT sequence if it is in the $\uparrow$ state, so that the resulting gate is actually CNOT$\otimes R_2 \!\left(\theta\right)$.  We can mitigate this by adding a $\text{CTAP}\!\!\downarrow \overline{\text{CTAP}\!\!\downarrow}$ pulse sequence at the end so that the control qubit picks up the same phase regardless of its state.  Then the effect of small constant $\Delta_1$ is just to add an \emph{overall} phase to the two-qubit state and the resulting gate is $e^{i \theta}$CNOT.  In Fig.~\ref{fig:pulseseq} we show an adiabatic pulse sequence to perform this gate.  In Fig.~\ref{fig:evolution} we show the corresponding time evolution of an initial state $|\Psi\left(0\right)\rangle = |3\rangle + 2|4\rangle -|10\rangle$ to $|\Psi\left(\tau\right) \rangle =|3\rangle - |4\rangle +2 |10\rangle$ for the ideal case of $\Delta_j = 0$, $J=0$, and constant $t_0$.  For the pulse widths shown, the gate has an error rate of $\epsilon_{\text{avg}} \sim 10^{-5}$.

(Note that we neglect loss of fidelity due to decoherence \citep{Greentree04} during the CTAP pulse sequence.  This is reasonable since the coherence time is typically more than four orders of magnitude longer than the CTAP pulse duration.  Likewise, the dots should be decoupled from any quantum point contact during the pulse to avoid loss of fidelity due to measurement backaction \citep{Rech11}.)
\begin{figure}
  \subfigure[]{\includegraphics[width=.98\columnwidth]{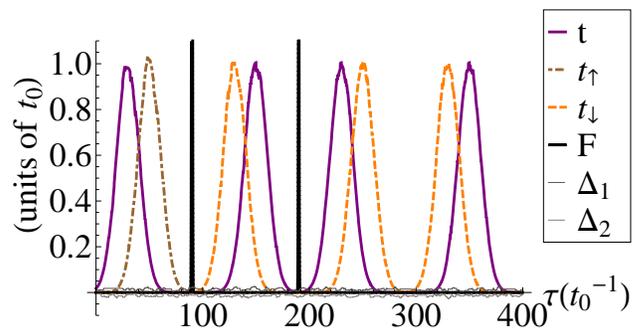}\label{fig:fastnoise}}
  \subfigure[]{\includegraphics[width=.98\columnwidth]{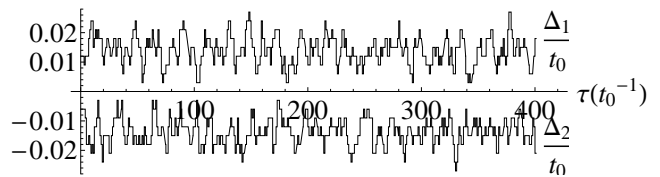}\label{fig:fastnoiseb}}
  \subfigure[]{\includegraphics[width=.98\columnwidth]{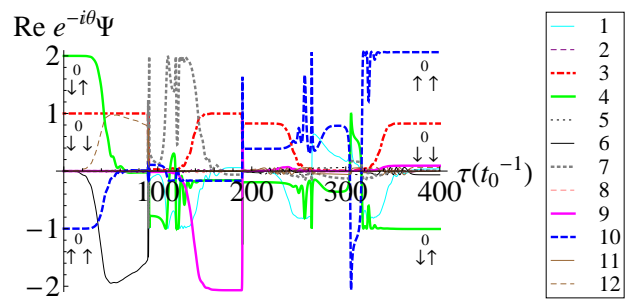}\label{fig:fastnoiseevolution}}\\
  \caption{(Color online.) (a) Fast noise in the parameters from 10 random telegraph impurities. (b) Close-up of detuning noise. (c) Real part of the evolution of initial state $|\Psi\left(0\right)\rangle =|3\rangle + 2|4\rangle -|10\rangle$ to $|\Psi\left(400/t_0\right)\rangle =|3\rangle - |4\rangle +2 |10\rangle$ under the CNOT pulse sequence for the noise shown in (a) and $J=t^2/200t_0$.  In this case the error is $\epsilon=5\times 10^{-3}$.}
\end{figure}

In a more realistic scenario, gate errors occur when charge noise results in large or fluctuating values of $\Delta_1$ during the procedure.  (Nonzero $\Delta_2$ has an effect only through the transitory population in the intermediate states of the CTAP processes, which can be made arbitrarily small by enhancing adiabaticity.)  We have checked numerically that, in the absence of tunneling noise, $\epsilon_{\text{avg}} < 10^{-4}$ for a constant $\Delta_1/t_0 < 0.1$.  For fluctuating $\Delta_1$, we consider three cases:

(1) If $\Delta_1$ fluctuates very quickly on the timescale of the CNOT operation, we have checked numerically that a sub-threshold error rate can be maintained if the magnitude of the fluctuations is $< 0.01 t_0$.  Figure \ref{fig:fastnoise} shows detuning fluctuations of $\sim 0.02 t_0$ and 2\% tunneling noise due to 10 random-telegraph-type charge impurities with a random assortment of capture and emission rates in the range $0.01-1 t_0$.  Figure \ref{fig:fastnoiseevolution} shows time evolution (removing the overall time-dependent phase discussed above) in the presence of that noise, with $J=t^2/200t_0$.  The error rate for this case is $\epsilon_{\text{avg}} \sim 10^{-2}$, which is worse than with a faster Loss-DiVincenzo gate.  So, in this case there is no advantage in using the CTAP pulse sequence.

(2) If $\Delta_1$ fluctuates on the timescale of the CNOT operation, the phases accumulated before and after the $\overline{\text{FLIP}}$ differ and we are back to having a CNOT$\otimes R_2 \!\left(\theta\right)$ gate with $\theta$ random.  The procedure is not robust against this type of noise: In the worst case scenario of a single jump near the middle of every CNOT operation the magnitude of the fluctuations clearly must be smaller than the gate frequency.  We have verified numerically for the pulse widths shown that the fluctuations must be $< 10^{-3} t_0$ to ensure $\epsilon_{\text{avg}} \sim 10^{-4}$.  Here again, the adiabatic pulse sequence is useless if detuning noise is appreciable.

(3) If $\Delta_1$ fluctuates very slowly on the timescale of the CNOT operation, the most likely source of error is a single jump of $\Delta_1$ during the procedure, as considered in case 2.  Here, though, by assumption, $\Delta_1$ does not jump during every CNOT operation, but only rarely, so the restriction of case 2 can easily be relaxed while maintaining a sub-threshold average error rate.  This case of low-frequency noise seems to be the most relevant for experiments \citep{Fujisawa00,Jung04,Petersson10,Hofheinz06}, and this is also where the adiabatic CNOT implementation is most robust.  Here errors essentially occur only when the parameters jump, not during the long intermediate times when the parameters take random quasi-static values.  In contrast, the exchange-coupled $\sqrt{\text{SWAP}}$ gate has a high error rate during the entire time the coupling is perturbed.

To summarize, competing requirements must be met for our proposal to be useful in a physical system with nonnegligible interdot detuning noise: adiabaticity, gate time fast compared to decoherence and noise switching time, and negligible exchange pulse area. Satisfying these conditions requires $T_2, T_{\text{RTN}} \gg 1/t_0 \gg 1/U$, where the inequalities are by several orders of magnitude.  The gate speed is limited by the largest achievable $U$, which is important in keeping the exchange coupling negligible.  For a maximal $U \sim 100$ meV, corresponding to a dot size $\sim$ nm, the resultant minimal gate time is $\tau \sim 10$ ns.

As an example of the possible utility of the adiabatic approach, we take the case of quantum dots in Si with maximal tunnel coupling $t_0 \sim \mu$eV.  Assume fluctuations in $t_0$ with a standard deviation of $10\%$.  The interdot detunings may also fluctuate on the $\mu$eV scale at predominantly low frequency.  Assume an average switching time on the order of seconds \citep{Hofheinz06}, so the effect of the charge noise is to introduce quasi-static uncertainties to the parameters.  If one implements an exchange-coupled $\sqrt{\text{SWAP}}$ gate with $U \sim 10\mu$eV, the gate time is $\sim 0.1\mu$s and the error rate is $\epsilon_{\text{avg}} \sim 10^{-2}$.  On the other hand, if one implements the proposed adiabatic CNOT gate with $U \sim$ meV and $\tau \sim \mu$s, the error rate decreases to $\epsilon_{\text{avg}} \sim 10^{-4}$.  The tradeoff is that the gate time (excluding the time required for the single-qubit flips) at a given tunneling rate has increased by an order of magnitude, but the error rate is suppressed by two orders of magnitude.

In AlGaAs/GaAs quantum dots, the shorter coherence time makes this tradeoff more important.  However, given the recent observation of $T_2 \sim 100\mu$s for a singlet-triplet qubit with dynamical decoupling \citep{coherence}, it is conceivable that similar coherence times could be achieved for single-spin qubits and $10^4$ adiabatic gate operations could be performed at $100$ MHz for $t_0 \sim 100\mu$eV and $U \sim 100$ meV.  At those high tunneling rates, a sub-threshold error rate persists for typical interdot detuning fluctuations of $\sim 0.1\mu$eV \citep{Jung04} even in the worst case scenario (2) where the noise is dominated by frequency components on the order of the gate frequency.

\section{Spin-dependent tunneling}
\begin{figure}
  \includegraphics[width=.5\columnwidth]{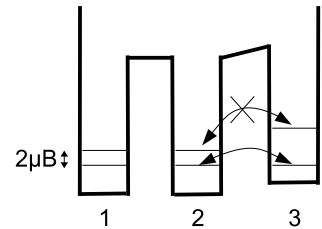}\\
  \caption{Possible spin-dependent tunneling scheme.}\label{fig:spindep}
\end{figure}

Obviously, a key requirement of our scheme is a controllable spin-dependent tunneling mechanism.  We briefly outline here one possible realization.  Suppose the Zeeman splitting on the auxiliary dot is made different from that on the qubit dots by engineering the Lande g-factor or placing a micromagnet nearby.  Then, by detuning the auxiliary and qubit dot potential wells relative to each other, the tunneling for a given spin state can be made resonant while for the other it is strongly suppressed for large detuning.  We sketch this scenario in Fig.~\ref{fig:spindep}.  This idea has previously appeared in the context of single-shot readout of a spin state \citep{Engel04}, and a closely related scheme with photon-assisted tunneling \citep{Barret06} has been experimentally demonstrated very recently \citep{Shin10}.  Note that the detuning of the auxiliary dot potential here in the presence of a magnetic field is \emph{not} the same as the $\Delta_j$ of the previous section, which corresponds to the energy mismatch of the energy levels relevant to the desired tunneling process and is zero in Fig.~\ref{fig:spindep}.

\section{Summary}
We have proposed an experimental procedure to perform a CNOT gate on two quantum dot spin qubits using an auxiliary third dot with controllable spin-dependent tunnel coupling.  The putative advantages of our proposed implementation are that it does not require precise control of the gate voltage pulse area or precise knowledge of the exchange coupling and thus it is robust against low-frequency noise.  For this method to work, several strict requirements must be met and, though rough estimates for Si and GaAs dots look promising, the immediate practicality of our protocol in these systems is far from certain and it is not clear whether this provides an easier path to fault-tolerant computation.  However, in light of the many candidate systems for exchange-based spin qubit quantum computation and the general nature of the proposal, we are hopeful that it will be applicable in at least one of them.  Also, materials development may make our scheme feasible for application to semiconductor quantum dot spin qubits in the future with the motivation coming from the acute need for suppressing charge noise induced decoherence in these qubits.

This work is supported by LPS-NSA and IARPA.


\begin{thebibliography}{99}
\bibitem{note} We time-evolve 100 initial states, randomly selected from the two-qubit subspace, under the pulse sequence and average over the resulting values of $\epsilon$ (which are all of the same order of magnitude).
\end{thebibliography}
\end{document}